\documentclass[10pt, twocolumn, superscriptaddress,preprintnumbers,showpacs,showkeys]{revtex4}
\usepackage{epsf} 
\usepackage{amsmath}
\usepackage{amssymb}
\usepackage{epsfig}
\usepackage{latexsym}
\usepackage{amsfonts}
\usepackage{graphicx}
\usepackage{varioref}
\usepackage{ifthen}
\begin{document}
\parindent 0mm 
\setlength{\parskip}{\baselineskip} 
\thispagestyle{empty}
\pagenumbering{arabic} 
\setcounter{page}{1}
\mbox{ }
\preprint{UCT-TP-299/14}

\title	{Scalar form factor of the pion in the Kroll-Lee-Zumino field theory}
\author{C. A. Dominguez}
\affiliation{Centre for Theoretical \& Mathematical Physics, and Department of Physics, University of
Cape Town, Rondebosch 7700, South Africa}
\author{M. Loewe}
\affiliation{Centre for Theoretical \& Mathematical Physics, and Department of Physics, University of
Cape Town, Rondebosch 7700, South Africa}
\affiliation{Instituto de F\'{\i}sica, Pontificia Universidad Cat\'{o}lica de Chile, Casilla 306, Santiago 22, Chile}
\author{M. Lushozi}
\affiliation{Centre for Theoretical \& Mathematical Physics, and Department of Physics, University of
Cape Town, Rondebosch 7700, South Africa}%

\date{\today}
\begin{abstract}
The renormalizable Kroll-Lee-Zumino field theory of pions and a neutral rho-meson is used to determine the scalar form factor of the pion in the space-like region at next-to-leading order. Perturbative calculations in this framework are parameter free, as the masses and the rho-pion-pion coupling are known from experiment. Results  compare favorably with lattice QCD calculations. \\
\end{abstract}

\maketitle
\noindent
The scalar form factor of the pion \cite{reviewFS}, and particularly its quadratic radius, plays an important role in chiral perturbation theory (CHPT) \cite{CHPT}. This form factor is defined as the pion matrix element of the QCD scalar current $J_S = m_u \bar{u}u + m_d \bar{d}d$, i.e.
\begin{equation}
F_S(q^2) = \langle \pi(p_2) | J_S | \pi(p_1) \rangle \;,
\end{equation}
where $q^2 = (p_2 - p_1)^2$. The associated quadratic scalar radius is given by
\begin{equation}
F_S(q^2) = F_S(0) \Bigl[ 1 + \frac{1}{6} \langle r^2_\pi\rangle_S \; q^2 +...\Bigr],
\end{equation}
where $F_S(0)$ is the pion sigma term
\begin{equation}
F_S(0) \equiv \sigma_\pi= m_q \frac{\partial M_\pi^2}{\partial m_q}.
\end{equation}
The scalar radius  fixes $\bar{ l }_4$, one of the low energy constants of CHPT, through the relation
\begin{equation}
\langle r^2_\pi\rangle_S = \frac{3}{8 \pi^2 F_\pi^2}\;\Bigl[ \bar{\ell}_4 - \frac{13}{12} +
O (M_\pi^2) \Bigr]\;,
\end{equation}
where $F_\pi = 91.9 \pm 0.1 \;\mbox{MeV}$ is the physical pion decay constant \cite{PDG}. The low energy constant $\bar{\ell}_4$, in turn, determines the leading contribution in the chiral expansion of the pion decay constant, i.e.
\begin{equation}
\frac{F_\pi}{F} = 1 + \left(\frac{M_\pi}{4 \pi F_\pi}\right)^2 \;\bar{\ell}_4 + O(M_\pi^4),
\end{equation}
where F is the pion decay constant in the chiral limit.\\
This scalar form factor is not accessible experimentally, but it has been determined from lattice QCD (LQCD) \cite{LQCD1}-\cite{LQCD2}, or hadronic models \cite{Hmodels}.\\

 Theoretically, the ideal tool to study this form factor, independently from LQCD, is the  Kroll-Lee-Zumino Abelian renormalizable field theory of pions and a neutral $\rho$-meson \cite{KLZ}. This  provides the appropriate field theory platform for the phenomenological Vector Meson Dominance (VMD) model \cite{VMD}, allowing for a systematic calculation of higher order quantum corrections \cite{CAD1}-\cite{CAD2}. Due to the renormalizability of the theory, predictions are parameter free, as the strong $\rho\pi\pi$ coupling, $g_{\rho\pi\pi}$, is known from experiment.
  In spite of this  coupling being a strong interaction quantity, perturbative calculations in the $\overline{MS}$ scheme make sense because the effective expansion parameter turns out to be $(g_{\rho\pi\pi}/4 \pi)^2 \simeq 0.2$.\\
     The KLZ theory has been used to compute the next-to-leading order (NLO) correction to the tree level (VMD) electromagnetic form factor of the pion in the space-like region with very good results \cite{CAD1}. In fact, it agrees with data up to $q^2 \simeq - 10 \; {\mbox{GeV}^2}$ with a chi-squared per degree of freedom  $\chi_F^2 = 1.1$, as opposed to VMD  which gives $\chi_F^2 = 5.0$. In addition, the mean-squared radius at NLO is $\langle r_\pi^2 \rangle = 0.46 \; {\mbox{fm}^2}$, compared with the experimental result \cite{PDG} $\langle r_\pi^2 \rangle = 0.45 \pm 0.01 \; {\mbox{fm}^2}$, and the  VMD value $\langle r_\pi^2 \rangle = 0.39 \; {\mbox{fm}^2}$.\\

\begin{figure}
[ht]
\begin{center}
\includegraphics[width=1.6 in]{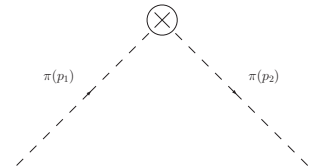}

\caption{\scriptsize{Leading order (LO) contribution to the scalar form factor of the pion. The cross indicates the coupling of the scalar current to two pions.}}
\end{center}
\end{figure}

\begin{figure}
[ht]
\begin{center}
\includegraphics[height=1.5 in,width=1.6 in]{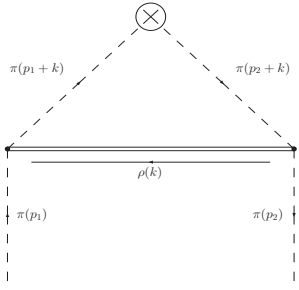}
\caption{Next-to-leading order (NLO) contribution to the scalar form factor.}
\end{center}
\end{figure}

In this note we compute in this framework the scalar form factor of the pion at NLO in the space-like region, and compare with current results from LQCD.\\
 The KLZ Lagrangian is given by

\begin{eqnarray}
\mathcal{L}_{KLZ} &=& \partial_\mu \phi \, \partial^\mu \phi^* -  M_\pi^2 \,\phi \,\phi^* - \tfrac{1}{4}\, \rho_{\mu\nu} \,\rho^{\mu\nu} 
+ \tfrac{1}{2}\, M_\rho^2\, \rho_\mu \,\rho^\mu \nonumber\\ [.3cm]
&+&g_{\rho\pi\pi} \rho_\mu J^\mu_\pi\ + g_{\rho\pi\pi}^2 \;\rho_\mu\; \rho^\mu \;\phi \;\phi^*   \;,
\end{eqnarray}

where $\rho_\mu$ is a vector field describing the $\rho^0$ meson ($\partial_\mu \rho^\mu = 0$), $\phi$ is a complex pseudo-scalar field describing the $\pi^\pm$ mesons, $\rho_{\mu\nu}$ is the usual field strength tensor: $\rho_{\mu\nu}  = \partial_\mu \rho_\nu - \partial_\nu \rho_\mu$, and $J^\mu_\pi$ is the $\pi^\pm$ current: $J^\mu_\pi  = i \phi^{*} \overleftrightarrow{\partial_\mu} \phi$. In spite of the explicit presence of the $\rho^0$ mass term in the Lagrangian, the theory is renormalizable because the neutral vector meson is coupled to a conserved current \cite{KLZ}. Figures 1 and 2 show, respectively, the LO and the NLO diagrams, where the cross indicates the coupling of the current to the two pions. Notice that while the Lagrangian, Eq.(6), contains a $\rho \rho \pi \pi$ quartic coupling, this term only contributes in this application at NNLO and beyond. \\

Using the Feynman propagator for the $\rho$-meson, and in $d$ dimensions, the unrenormalized vertex function in Fig.2 in dimensional regularization  is given by 
\begin{eqnarray}
\widetilde{G} (q^2) &=&  -2\;\frac{g_{\rho\pi\pi}^2}{(4 \pi)^2} \left( \mu^2 \right)^{(2 - \frac{d}{2})} \int_0^1 dx_1\int_0^{1-x_1} dx_2 \nonumber \\ [.3cm]
&\times&
\left\{ \frac{2}{\varepsilon} - \ln \left( \frac{\Delta(q^2)}{\mu^2}\right) 
- \frac{1}{2} - \gamma + \ln (4 \pi)  \right. \nonumber \\ [.3cm]
&+& \left. \frac{1}{2 \Delta(q^2)} \left[M_\pi^2 (x_1 + x_2 - 2)^2 
\right. \right.\nonumber \\ [.3cm]
&-& \left. \left. q^2 (x_1 x_2 - x_1 - x_2 +2)\right] + O(\varepsilon) \phantom{\frac{1}{1}} \right\} \;,
\end{eqnarray}

where $\Delta(q^2)$ is defined as
\begin{equation}
\Delta(q^2) = M_\pi^2 (x_1+x_2)^2 + M_\rho^2(1-x_1-x_2) - x_1 x_2 q^2 \;.
\end{equation}

In the $\overline{MS}$ scheme, and renormalizing the vertex function at the point $q^2=0$, the NLO contribution in Fig. 2 is  \cite{CAD2}

\begin{eqnarray}
&&G(q^2) - G(0) = - 2\;  \frac{g_{\rho\pi\pi}^2}{(4 \pi)^2} \int_0^1 dx_1 \int_0^{1-x_1} dx_2  \nonumber \\ [.3cm]
&\times& \left\{ \ln \left( \frac{\Delta(q^2)}{\Delta(0)}\right)  + \frac{1}{2}  \left[ M_\pi^2 (x_1 + x_2 - 2)^2 \left(\frac{1}{\Delta(q^2)}
\right. \right. \right. \nonumber \\ [.3cm]
&-&  \left. \left.  \left.  \frac{1}{\Delta(0)}\right)
- \frac{q^2}{\Delta(q^2)} (x_1 x_2 - x_1 - x_2 +2) \right] \right\} \;,
\end{eqnarray}

with

\begin{equation}
F_S(q^2) = F_S(0) \left[ 1 + G(q^2) - G(0)\right] \;.
\end{equation}

\begin{figure}
\begin{center}
\includegraphics[height=3.2 in,width=3.64 in]{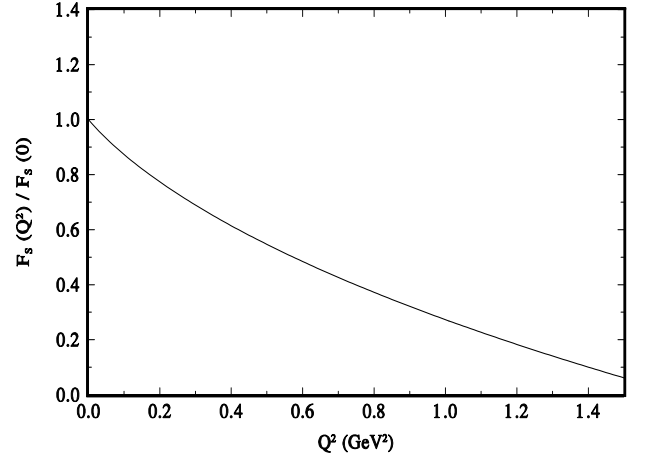}
\caption{The normalized scalar form factor, Eq.(8), to NLO in the space-like region.}
\end{center}
\end{figure}

For details on the renormalization procedure for the fields, masses and coupling see \cite{CAD1}.
The result of a numerical evaluation of Eq.(9), using $g^2_{\rho\pi\pi} = 36.0 \pm 0.2$ from the measured width of the $\rho$-meson \cite{PDG}, is shown in Fig.3. Regarding the scalar radius, defined in Eq.(2), we confirm the NLO result obtained in \cite{CAD2}
\begin{equation}
\langle r^2_\pi\rangle_S  = 0.4 \;{\mbox{fm}}^2 \;,
\end{equation}

with a negligible error due to the strong coupling.

This value is smaller than typical values in the literature \cite{LQCD1}-\cite{Hmodels}. However, it must be kept in mind that the NLO result is expected to be a lower bound, i.e. with $[G(q^2) - G(0)] < 0$ the NNLO would reduce $F_S(q^2)$, thus increasing the radius. A rough order of magnitude estimate of the size of the NNLO contribution suggests a correction of some 20\% to the NLO term (the NNLO calculation is quite formidable and beyond the scope of this note). This is obtained by estimating a typical two-loop diagram, e.g. the $\rho$-meson propagator at NNLO and comparing it with the NLO result. The Feynman integrals in the variables $x_i$ at NLO and NNLO are of order ${\cal{O}}(1)$ in the $q^2$ range explored here. We find the total contribution from this diagram to be over 20\% of the NLO, thus increasing the radius to $\langle r^2_\pi\rangle_S  \simeq 0.5 \;{\mbox{fm}}^2 $.\\

A comparison of the KLZ form factor itself at low $|q^2| < 0.5 \;{\mbox{GeV}^2}$ with LQCD results read from figures in \cite{LQCD1} and \cite{LQCD2} shows  good agreement.
It should be mentioned, though, that LQCD results from \cite{LQCD1} are for light-quark masses in the range from $m_s/6$ to $m_s/2$, while those from \cite{LQCD2} are for $m_\pi = 325 \; {\mbox{MeV}}$. These LQCD determinations find  values for the scalar radius  higher than in this analysis, Eq.(11), i.e. $\langle r^2_\pi\rangle_S  = 0.6 \pm 0.1 \;{\mbox{fm}}^2$ from \cite{LQCD1}, and  $\langle r^2_\pi\rangle_S  = 0.637 \pm 0.023 \;{\mbox{fm}}^2$ from \cite{LQCD2}. These results for the radius are determined from e.g. chiral extrapolations to the physical pion mass. Our results for the form factor are also in agreement within less than 10\% with a CHPT calculation \cite{CHPT2} in the range $-q^2 = 0 - 0.2 \; {\mbox{GeV}^2}$.

{\bf{Acknowledgements}}
This work was supported in part by FONDECyT (Chile) under grants 1130056 and 1120770, by NRF (South Africa), and by the University of Cape Town URC. The authors wish to thank Gary Tupper for valuable discussions on KLZ, and Hartmut Wittig for enlightening exchanges on LQCD.

\end{document}